\documentstyle[apj,times,psfig]{article}

\def\ie{i.e$.$~} \def\eg{e.g$.$~} \def\eq{eq$.$~}  \def\etal{et al$.$~}
\def\tildelta{\tilde{\delta}} \def\cm3{{\rm cm}^{-3}}
\def\simg{\mathrel{%
      \rlap{\raise 0.511ex \hbox{$>$}}{\lower 0.511ex \hbox{$\sim$}}}}
\def\siml{\mathrel{%
      \rlap{\raise 0.511ex \hbox{$<$}}{\lower 0.511ex \hbox{$\sim$}}}}

\begin{document}

\title{The Effect of Angular Structure of Gamma-Ray Burst Outflows on the Afterglow Emission}

\author{A. Panaitescu and P. Kumar}
\affil{Department of Astronomy, University of Texas, Austin, TX 78712}

\begin{abstract}
 
 We investigate analytically the effect of an anisotropic distribution of the kinetic energy
within a relativistic fireball on the decay of the afterglow emission, focusing on axially 
symmetric fireballs with a uniform core and a power-law decrease with angle outside the core. 
The afterglow fall-off steepens after the core becomes fully visible. For observer directions 
within the core, simple formulae are derived for the afterglow decay after the break, while for 
off-core observers results are shown graphically. Some criteria for assessing from observations
the necessity of an angular structure and/or collimation are given. Applying them to several 
afterglows with light-curve breaks, we find that jets endowed with structure are required only 
if the circumburst medium has a wind-like ($r^{-2}$) stratification. Numerical fits to the 
multiwavelength emission of the afterglows 990510 and 000301c show that, for the former, the 
angular distribution of the ejecta kinetic energy is not far from uniformity, and that, with 
a standard power-law electron distribution, the sharp steepening of the $R$-band light-curve 
of the latter is better accommodated by a structured jet than an uniform outflow. Structured
outflows may accommodate the shallower light-curve breaks observed in other afterglows.

\end{abstract}

\keywords{gamma-rays: bursts - ISM: jets and outflows - methods: numerical -
          radiation mechanisms: non-thermal - shock waves}

\section{Introduction}

 Almost all the analysis of the decaying light-curves of Gamma-Ray Burst (GRB) afterglows 
is done within the framework of external shocks driven into the circumburst medium by 
ultrarelativistic ejecta (M\'esz\'aros \& Rees 1997) whose kinetic energy is the same in 
all directions
\footnote {Throughout this article, we use the term "uniform" to designate such outflows.
If the angular distribution of the energy is not isotropic, the outflow will be called
"structured". Tightly collimated outflows undergoing a significant lateral spreading 
at the time of observations will be referred to as "jets", while "fireballs" will be 
used if the lateral spreading is negligible.}.
A non-isotropic distribution of the energy per solid angle within the outflow is a natural 
extension of the afterglow model. Fireballs whose kinetic energy and initial Lorentz factor 
fall-off with angle are power-laws have been considered for the first time by M\'esz\'aros, 
Rees \& Wijers (1998), who studied the effect of such distributions on the afterglow 
light-curve decay. The faster dimming of the afterglow emission that a structured outflow
can produce has been used by Dai \& Gou (2001) to explain the steep fall-off of the optical
light-curve of the afterglow 991208.  Postnov, Prokhorov and Lipunov (2001) have suggested 
that GRB outflows may have a universal angular structure, the observed distribution of 
isotropic $\gamma$-ray outputs (which has a width of 3 dex) being due to the observer location. 
Rossi, Lazzati \& Rees (2002) and Zhang \& M\'esz\'aros (2002) have proposed that the 
light-curve breaks seen in several GRB afterglows and the narrow distributions of the GRB 
energy release (Frail \etal 2001) and jet kinetic energy (Panaitescu \& Kumar 2002) may be 
due to the angular structure of fireballs.

 In this work, we present an analytical treatment of the afterglow light-curves from structured
fireballs (\S\ref{analytical}), focusing on axially symmetric fireballs endowed with a power-law 
distribution of the energy (\S\ref{plaw}). In \S\ref{criteria} we give criteria which can be used
to assess from the afterglow properties when structure and collimation of GRB outflows is required
by observations, and apply these criteria to the afterglows whose optical light-curves exhibited 
a break. Section \S\ref{models} presents the numerical modeling of two GRB afterglows, 990510 and 
000301c, in the framework of structured jets, leading to a few important conclusions about the 
role of structure in these two cases.

\section{Synchrotron Emission from Structured Fireballs}
\label{analytical}
\vspace*{2mm}

  In this section we calculate the evolution of the afterglow light-curve index $\alpha (t)$, 
defined as the logarithmic derivative with respect to the observer time $t$ of the received 
flux $F_\nu$ (\ie $F_\nu \propto t^{-\alpha}$) for a fireball endowed with structure. 
Our aim is to obtain the dependence of of $\alpha (t)$ and its asymptotic values at early and 
late times on the fireball's angular structure, \ie the sharpness of the afterglow light-curve 
break that the structure can produce.
 
  The calculation of the afterglow light-curve requires the following ingredients: dynamics of 
the fireball, spectrum of its emission, and integration over the equal photon-arrival-time surface. 
For analytical calculations, we shall ignore the tangential motions and mixing in the fireball
and consider a simplified scenario where a fluid patch travels as if it were part of a uniform 
fireball.

 For simplicity, in our analytical calculations of the light-curve index we consider only adiabatic
GRB remnants, a case encountered when electrons acquire a negligible fraction of the post-shock
energy or if they cool on a timescale longer than the dynamical time. Radiative losses could be
important during the early afterglow, in which case the results presented in this section should
be re-derived taking into account their effect on the evolution of the fireball
Lorentz factor $\gamma$ [\eq (\ref{gm})]. This effect is included in the numerical calculations 
presented in \S\ref{0510} and \S\ref{0301}.

 For an adiabatic fireball, energy conservation gives that the Lorentz factor of a fluid patch
moving in the direction ($\delta, \psi$) ($\delta$ and $\psi$ being the polar and azimuthal 
angles in a spherical polar coordinate system aligned to the direction toward the observer)
decreases with its radial location $r$ according to: 
\begin{equation}
 \gamma = K [{\cal E}(\delta,\psi)]^{1/2}\, r^{-(3-s)/2} \;,
\label{gm}
\end{equation}
where $K$ is a constant, independent of direction, ${\cal E}$ is the kinetic energy per solid
angle of the ejecta moving in the  ($\delta, \psi$) direction, and $s$ describes the type of
external medium: $n(r)\propto r^{-s}$, $s=0$ for a homogeneous medium and $s=2$ for a wind
ejected at constant speed and mass-loss rate. Equation (\ref{gm}) holds for $r$ larger than 
the radius at which the fireball deceleration sets in and until the fireball becomes 
semi-relativistic. Hereafter we will focus on this case, as it applies to observer times 
of at least 10 days, when the afterglow light-curve breaks are seen.

 The photons emitted by a patch ($d\delta, d\psi$) arrive at observer at a time $t$ given by
integrating
\begin{equation}
  dt = dr (1-\beta \cos \delta) \;,
\label{tobs}
\end{equation}
where $\beta$ is the velocity of the patch, and the speed of light is unity.
In the limit of relativistic motion ($\gamma \gg 1$) most radiation received by observer comes
from the shocked medium moving at small angles ($\delta \ll 1$) relative to the line of sight.
With these two assumptions, equation (\ref{tobs}) leads to the following equation for the
radius $r$ of the patch moving in direction ($\delta,\psi$) as a function of observer time
\begin{equation}
  \frac{1}{2} r \delta^2 + \frac{1}{2(4-s)} \frac{r^{4-s}}{K^2 {\cal E}} = t \;.
\label{time}
\end{equation}

 The peak synchrotron flux $F(\nu_p)$ received from a $(d \delta, d \psi)$ patch is proportional
to ${\cal D}^3 I'(\nu'_p) d\Sigma$, where 
\begin{equation}
 {\cal D} = \frac{1}{\gamma (1-\beta\cos\delta)} \simeq \frac{2 \gamma}{1+ \gamma^2 \delta^2} \;
\label{doppler}
\end{equation} 
is the Doppler boost factor, $I'(\nu'_p)$ is the comoving frame peak intensity at frequency 
$\nu'_p = \nu/{\cal D}(\delta)$ and $d\Sigma = r^2 \sin \delta\, d\delta\, d\psi$ is the area of 
the patch. The $I'(\nu'_p)$ is proportional to the product ${\cal N} B$ of the surface density
of energized electrons and the magnetic field within the shocked external medium. Therefore,
the flux received at some frequency $\nu$ above the self-absorption frequency is
\begin{equation}
  F_\nu (t) \propto \int_{\Omega} \frac{\gamma^3 {\cal N} B {\cal F}(\nu)}
                 {(1 + \gamma^2 \delta ^2)^3} \; \delta\, d\delta\, d\psi \; 
\label{fnu1}
\end{equation}
where the integral is over all directions of ejecta motion, 
\begin{equation}
 {\cal N} \propto r^{-2} \int_0^r n(r')r'^2 dr' \propto r^{1-s} \;,
\label{N}
\end{equation}
\begin{equation}
 B^2 \propto n \gamma^2 \propto r^{-s} \gamma^2 \;,
\label{Bmag}
\end{equation}
assuming a magnetic energy density that is a constant fraction of the post-shock energy density,
(which is the product the typical random Lorentz factor acquired by protons, $\gamma$, and the 
comoving proton density, $4 n \gamma$), and the factor ${\cal F}(\nu)$ is the ratio of the flux 
at frequency $\nu$ to that at the peak $\nu_p$. The time dependence of the flux given in equation
(\ref{fnu1}) is due to the variation of $\gamma$, $\cal N$, and $B$ with the location $r$ of the 
patch, which is related to the observer time through equation (\ref{time}).
By substituting $\cal N$ and $B$ in equation (\ref{fnu1}), the flux becomes
\begin{equation}
 F_\nu(t) \propto \int_\Omega r^{-3+0.5s}\, (1+\gamma^2 \delta^2)^{-3}\, {\cal E} ^2 
                           {\cal F}(\nu)\;  \delta\, d\delta\, d\psi \;.
\label{fnu2}
\end{equation}

 The spectral factor ${\cal F} (\nu)$ is determined by the distribution of the radiating 
electrons. This distribution is shaped by the continuous injection in the downstream region of 
shock-accelerated electrons with a power-law distribution in energy, $dN_i/d\epsilon \propto 
\epsilon^{-p}$, above a minimum electron energy $\epsilon_i \propto \gamma$, and by radiative 
cooling. The resulting electron distribution exhibits a cooling break at an energy $\epsilon_c$ 
for which the cooling timescale ($\propto (\epsilon_c B^2)^{-1}$) equals the age of the fireball
($\propto r/\gamma$), therefore $\epsilon_c \propto \gamma/(rB^2)$.
Above $\epsilon_c$ the effective electron distribution is $dN/d\epsilon  \propto \epsilon^{-(p+1)}$ 
if $\epsilon_i < \epsilon_c$ (\eg Kardashev 1962) or $dN/d\epsilon \propto \epsilon^{-2}$ if
$\epsilon_c < \epsilon_i$. Consequently, the synchrotron spectrum is piecewise power-law, with 
breaks at the injection frequency $\nu_i \propto {\cal D} \epsilon_i^2 B$ and cooling frequency 
$\nu_c \propto {\cal D} \epsilon_c^2 B$:
\begin{equation}
  {\cal F}(\nu) = \left\{ \hspace*{-2mm} \begin{array}{ll}   
       (\nu/\nu_i)^{1/3}                       &  \nu < \nu_i           \\
       (\nu/\nu_i)^{-(p-1)/2}                  &  \nu_i < \nu < \nu_c   \\
         \end{array} \right. \;,
\label{spek1}
\end{equation}
for $\epsilon_i > \epsilon_c$, while for $\epsilon_c < \epsilon_i$
\begin{equation}
  {\cal F}(\nu) = \left\{ \hspace*{-2mm} \begin{array}{ll}   
       (\nu/\nu_c)^{1/3}                       &  \nu < \nu_c           \\
       (\nu/\nu_c)^{-1/2}                      &  \nu_c < \nu < \nu_i   \\
         \end{array} \right. \;.
\label{spek2}
\end{equation}
If $\nu$ is above both spectral breaks, then
\begin{equation}
  {\cal F}(\nu) =  (\nu/\nu_i)^{-p/2} (\nu_c/\nu_i)^{1/2} \;,
\label{spek3}
\end{equation}
for either ordering of $\nu_i$ and $\nu_c$. Thus, the time dependence of the factor ${\cal F}(\nu)$ 
is determined by those of the spectral breaks. After using equations (\ref{gm}) and (\ref{Bmag}),
one obtains:
\begin{equation}
 \nu_i \propto  \frac{r^{-6+1.5 s} {\cal E}^2}{1+\gamma^2 \delta^2} \;,\;
  \frac{\nu_c}{\nu_i} \propto \frac{r^4}{{\cal E}^2} \;.
\label{nu}
\end{equation}

 For a given fireball structure ${\cal E}(\delta,\psi)$, one can calculate the afterglow light-curve
$F_\nu (t)$ by integrating numerically equation (\ref{fnu2}) with $r$, $\gamma$, and ${\cal F}(\nu)$ 
corresponding to arrival time $t$ and direction $(\delta,\psi)$ as given by equations (\ref{time}), 
(\ref{gm}), (\ref{spek1})--(\ref{spek3}), and (\ref{nu}). 
For now we simplify the integral in equation (\ref{fnu2}) by using an approximate solution to equation 
(\ref{time}). In that equation, the second term in the left hand side is dominant for small $\delta$,
while the first term dominates at large $\delta$. The two terms become equal for $r(\delta,\psi)=
[(4-s)K^2 \delta {\cal E}]$, which describes a curve $(\tildelta (\psi),\psi)$ on the equal photon 
arrival-time surface of time $t$ given by  
\begin{equation}
 {\cal E}(\tildelta, \psi) \tildelta^{2(4-s)} = \frac{t^{3-s}}{(4-s)K^2} \;.
\label{tildelta}
\end{equation}
From equation (\ref{gm}), the above expression for $r$ shows that $\tildelta(\psi)$ also satisfies 
$\tildelta(\psi) \, \gamma (\tildelta,\psi) = (4-s)^{-1/2}$.

 For $\delta < \tildelta(\psi)$, equations (\ref{gm}) and (\ref{time}) give
\begin{equation}
 r \propto ({\cal E} t)^{1/(4-s)} \;,\;\; \gamma \propto ({\cal E} t^{s-3})^{1/2(4-s)} \;,
\label{rg1}
\end{equation}
while for $\tildelta(\psi) < \delta$ 
\begin{equation}
 r \propto t/\delta^2 \;,\;\; \gamma \propto \delta^{3-s} ({\cal E} t^{s-3})^{1/2} \;.
\label{rg2}
\end{equation}

 The integral in equation (\ref{fnu2}) can then be approximated as the sum of an ``inner" 
integral over the $\delta < \tildelta (\psi)$ disk, where $r$ and $\gamma$ are given by equation 
(\ref{rg1}), and an ``outer" integral over directions outside this disk, where $r$ and $\gamma$ 
are given by equation (\ref{rg2}). It can be shown by numerical integration that these 
two integrals have the same time dependence, thus the calculation of the afterglow light-curve
can be simplified by retaining only the inner integral. The product $\gamma \delta$ increases 
with $\delta$, thus its maximal value on the inner disk is $\tildelta \gamma (\tildelta,\psi) = 
(4-s)^{-1/2} < 1$, therefore the factor $(1+\gamma^2 \delta^2)^{-3}$ in equation (\ref{fnu2}) 
is approximately unity. With these simplifications, the afterglow flux becomes:
\begin{equation}
 F_\nu \propto t^{(0.5s-3)/(4-s)} \int\limits_0^{2\pi} d\psi \int\limits_0^{\tildelta(t,\psi)} 
               d\delta\; \delta\, {\cal E}^{(5-1.5s)/(4-s)}\, {\cal F} (\nu) \;,
\label{fnu3}
\end{equation}
where ${\cal F} (\nu)$ is given by equations (\ref{spek1})--(\ref{spek3}) with 
\begin{equation}
  \frac{\nu}{\nu_i} \propto {\cal E}^{-1/2} t^{3/2} \;,\;
  \frac{\nu_c}{\nu_i} \propto \left( {\cal E}^{s-2} t^2 \right)^{2/(4-s)} \;.
\label{Nu}
\end{equation}
Therefore the afterglow light-curve has the form
\begin{equation}
 F_\nu (t) \propto t^{f(p,s)} \int\limits_0^{2\pi} d\psi \int\limits_0^{\tildelta(t,\psi)} 
                   d\delta\; \delta\, [{\cal E}(\delta,\psi)]^{g(p,s)} \;,
\label{fnu4}
\end{equation}
where the exponents $f$ and $g$ depend on the electron index $p$ and type of external medium.

\vspace*{3mm}
\section{Power-Law Fireballs}
\label{plaw}
\vspace*{2mm}

 Hereafter we shall consider fireballs whose kinetic energy per solid angle ${\cal E}$ is
axially symmetric, uniform within a core of size $\theta_c$, and a power-law outside it
\footnote{Other types of structure can be considered as well, but power-laws allow sufficient 
          flexibility with a minimal number of new parameters} 
:
\begin{equation}
 {\cal E} (\theta) = {\cal E}_0 \times \left\{  \hspace*{-2mm} \begin{array}{ll}
                 1                 &  \theta < \theta_c   \\ 
          (\theta/\theta_c)^{-q}   &  \theta_c < \theta
       \end{array} \right. \;,
\label{dEdO}
\end{equation} 
where $q > 0$ and the angle $\theta$ measured from the symmetry axis of the fireball is related 
to the direction ($\delta,\psi$) through a rotation:
\begin{equation}
 \cos\theta = \cos\delta \, \cos\theta_{obs} - \sin\delta \, \cos\psi \, \sin\theta_{obs} \;, 
\end{equation}
$\theta_{obs}$ being the angle between the observer's line toward the fireball center and the
fireball symmetry axis. 

\vspace*{2mm}
\subsection{Observers within the Fireball Core: Analytical Results}
\vspace*{2mm}

 As shown by Granot \etal (2002), the emission from the uniform core depends very 
weakly on the observer location. The emission received from outside the fireball core 
should depend even less on the observer's off-set $\theta_{obs}$, thus, to a good 
approximation, we can consider $\theta_{obs}=0$. Then equation (\ref{tildelta}) gives
\begin{equation}
  \tildelta (t) \propto \left\{  \hspace*{-2mm} \begin{array}{ll} 
             t^{(3-s)/2(4-s)}      & {\rm if} \;\; \tildelta < \theta_c \\  
             t^{(3-s)/[2(4-s)-q]}  & {\rm if} \;\; \theta_c < \tildelta
       \end{array} \right. \;.
\end{equation}

 At early times, when $\tildelta < \theta_c$, the afterglow emission 
arises from within the uniform fireball core. In this case, the afterglow light-curve
is $F_\nu \propto t^{f(p,s)} \int_0^{\tildelta(t)} d\delta\, \delta \propto  t^{f(p,s)}
\tildelta^2 $, which leads to $F_\nu \propto t^{-\alpha_1}$ with
\begin{equation}
  \alpha_1 = \left\{\hspace*{-2mm} \begin{array}{ll} 
       (3p-3)/4  &  s=0   \;\; \& \;\;      \nu_i < \nu < \nu_c \\ 
       (3p-2)/4  &  s=0,2 \;\; \& \;\; \max\{\nu_i,\nu_c\} < \nu   \\ 
       (3p-1)/4  &  s=2   \;\; \& \;\;      \nu_i < \nu < \nu_c 
        \end{array} \right. \;.
\label{a1}
\end{equation}
These are the usual results for a uniform fireball. 

 At late times, when $\tildelta > \theta_c$, the contribution of the ejecta outside the 
core to the afterglow emission is proportional to 
\begin{equation}
 \int_{\theta_c}^{\tildelta (t)} \delta^{1-q g(p,s)} d\delta = 
         (2-q g)^{-1} \left(\tildelta^{2-qg} -\theta_c^{2-qg} \right) \;.
\end{equation}
If $q g(p,s) < 2$, then the first term in the right hand side dominates, while for
$q g(p,s) > 2$, the last term is the largest and has the same time dependence as the 
emission from the fireball core. Using the expression of $g(p,s)$ for each case,
the transition value of the structural parameter is 
\begin{equation}
  \tilde{q} = \left\{  \hspace*{-2mm} \begin{array}{ll}
    \frac{\displaystyle8}{\displaystyle p+4} & (s=0, \nu_i < \nu < \nu_c) \;\; 
                                    {\rm or} \;\; (s=2, \max\{\nu_i,\nu_c\} < \nu) \\ 
    \frac{\displaystyle 8}{\displaystyle p+3} & (s=0, \max\{\nu_i,\nu_c\} < \nu) \;\; 
                  {\rm or} \;\; (s=2, \nu_i < \nu < \nu_c)  
        \end{array} \right. \;.
\label{tildeq}
\end{equation}
For $q < \tilde{q}$, the fall-off of ${\cal E}(\theta)$ is sufficiently shallow that the 
emission from the ejecta outside the core is dominant, the afterglow light-curve 
$F_\nu \propto t^{f(p,s)} \tildelta^{2-qg}$ decaying as $F_\nu \propto t^{-\alpha_2}$ with
\begin{equation}
 (s=0) \quad \alpha_2 = \frac{1}{4-\frac{1}{2}q} \times \left\{  \hspace*{-2mm} 
       \begin{array}{ll} 
            3p-3+\frac{3}{2}q  &  \nu_i < \nu < \nu_c \\ 
            3p-2+q             &  \max\{\nu_i, \nu_c\} < \nu 
        \end{array} \right. \;,
\label{a20}
\end{equation}
for a homogeneous medium and
\begin{equation}
 (s=2) \;\; \alpha_2 = \frac{1}{4-q} \times \left\{  \hspace*{-2mm} \begin{array}{ll} 
           3p-1-\frac{1}{2}q(p-1)  &  \nu_i < \nu < \nu_c \\ 
           3p-2-\frac{1}{2}q(p-2)  &  \max\{\nu_i, \nu_c\} < \nu 
        \end{array} \right. \;,
\label{a22}
\end{equation}
for a wind-like medium.
For $q > \tilde{q}$, ${\cal E}(\theta)$ decreases sufficiently fast away from the fireball 
axis that the afterglow emission is dominated by the uniform core and/or by the ejecta just
outside the core. In either case, the afterglow light-curve $F_\nu \propto t^{f(p,s)} 
\theta_c^{2-qg}$ decays as  $F_\nu \propto t^{-\alpha_2}$ with
\begin{equation}
  \alpha_2 = \left\{\hspace*{-2mm} \begin{array}{ll} 
    3p/4     &  (s=0, \nu_i < \nu < \nu_c) \;\; {\rm or} \;\; (s=2, \max\{\nu_i,\nu_c\} < \nu) \\ 
    \frac{\displaystyle 3p+1}{\displaystyle 4} &  (s=0, \max\{\nu_i,\nu_c\} < \nu) \;\; 
                                                  {\rm or} \;\; (s=2, \nu_i < \nu < \nu_c)  
    \end{array} \right. \;.
\label{a2}
\end{equation}

 Taking into account that for most well-observed afterglows, the electron index is between 1.5 
and 2.5, it follows that $5/4 \siml \tilde{q} < 7/4$. Therefore, for fireballs that do not
undergo significant lateral spreading at the observing time, the asymptotic index $\alpha_2$ 
of the afterglow light-curve is independent of the fireball power-law structure if $q \simg 2$. 
This is true also for observers located outside the fireball core (\S\ref{offcore}), at times
when the core is visible. 
\footnote{A decelerating, point-like source moving at Lorentz factor $\gamma$ and at an angle 
   $\delta$ relative to the direction toward the observer, becomes visible when $\gamma \delta = 1$, 
   \ie when the cone of its relativistically beamed emission starts to contain the direction toward
   the observer}

 Equations (\ref{a1}) and (\ref{a20})--(\ref{a2}) allow the calculation of the steepening
of the light-curve index $\Delta \alpha = \alpha_2 - \alpha_1$ produced by the passage of 
the cooling frequency $\nu_c$ through the observing band or by the core edge becoming visible.
For the former, $\Delta \alpha = 1/4$ either if the entire visible region of the fireball is 
within its core or if the core edge is observable but $q > \tilde{q}$. For $q < \tilde{q}$, 
the passage of $\nu_c$ (which decreases in time for a homogeneous medium, but increases if 
the medium is wind-like) yields $\Delta \alpha = (2-q)/(8-q) < 1/4$. 

 Because $\gamma \delta$ increases with $\delta$ and $\tildelta \, \gamma (\tildelta) = (4-s)^{-1/2} 
< 1$, when the core edge becomes visible ($\gamma (t_c,\theta_c) \theta_c = 1$) the core edge
is outside the inner disk ($\theta_c > \tildelta (t_c)$). Therefore $\gamma (t_c,\theta_c)$ 
satisfies equation (\ref{rg2}), which, together with $\gamma (t_c,\theta_c) = \theta_c^{-1}$, 
yields
\begin{equation}
  t_c \propto [{\cal E}_0 \theta_c^{2(4-s)}]^{1/(3-s)} \;.
\label{tc}
\end{equation}
For $q > \tilde {q}$ the light-curve break across $t_c$ is $\Delta \alpha = 3/4$ for a 
homogeneous medium and $\Delta \alpha = 1/2$ for a wind, while for $q < \tilde{q}$ the break is 
\begin{equation}
 \Delta \alpha = \left\{ \begin{array}{ll} 
      \hspace*{-2mm}  \frac{\displaystyle 3q(8/\tilde{q}-1)}{\displaystyle 4(8-q)} < 3/4 & s=0  \\
      \hspace*{-2mm}  \frac{\displaystyle  q(4/\tilde{q}-1)}{\displaystyle 2(4-q)}  < 1/2 & s=2 
     \end{array} \right. \;.
\end{equation}
Therefore the maximal light-curve break that a non-spreading, power-law fireball can yield to 
an observer located with its uniform core is $3/4$ for a homogeneous medium and $1/2$ for a wind
(Panaitescu, M\'esz\'aros \& Rees 1998).

\vspace*{2mm}
\subsection{Observers within the Fireball Core: Numerical Results}
\vspace*{2mm}

 The evolution of the light-curve index $\alpha (t)$, obtained by calculating numerically the
integral of equation (\ref{fnu2}) (\ie without the simplifications made subsequently), for an 
on-axis observer is shown in Figure 1 for some values of the structural parameter $q$.
Before the edge core becomes visible, the index $\alpha = \alpha_1$ given by equation (\ref{a1}). 
When the core edge becomes visible, at $t=t_c \sim 2\times 10^3\, t_d$ for $s=0$ and $\sim 4 
\times 10^3\, t_d$ for $s=2$ ($t_d$ being the observer frame deceleration timescale for the fluid 
moving toward the observer: $\theta=0$), the index $\alpha$ increases toward the $\alpha_2$ 
given in equations (\ref{a20}) and (\ref{a22}) for $q = 1 < \tilde{q}$ and that given in equation 
(\ref{a2}) for $q = 2, 3 > \tilde{q}$. As shown in Figure 1, 80\% of the light-curve decay 
steepening across $t_c$ lasts a factor 4--6 in time for $s=0$ and a factor 7--10 for $s=2$. 
The slower transition for a wind-like medium is caused by the slower decrease of the fireball
Lorentz factor with time, given in equations (\ref{rg1}) and (\ref{rg2}): for $s=0$, $\gamma 
\propto t^{-3/8}$ if $\delta < \tildelta$ and $\gamma \propto t^{-3/2}$ if $\delta > \tildelta$,
while for $s=2$, $\gamma \propto t^{-1/4}$ and $\gamma \propto t^{-1/2}$, respectively. 
The fireball deceleration being slower for a wind-like medium, it takes a longer time for 
the core edge to become fully visible, thus the transition between the asymptotic light-curve
indices is smoother.

\begin{figure*}
\centerline{\psfig{figure=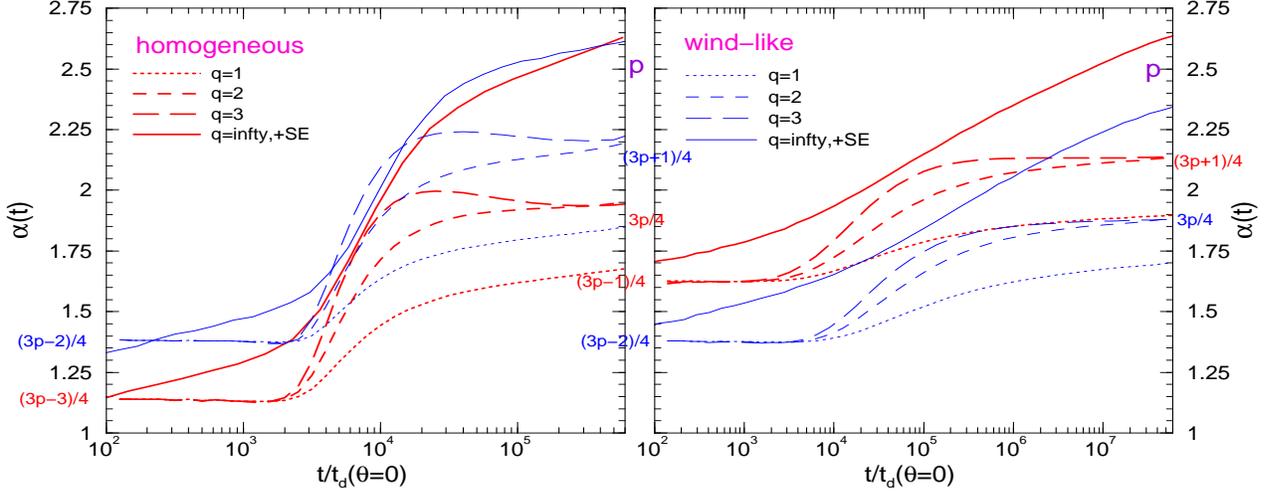,width=17cm,height=7cm}}
\figcaption{Evolution of the logarithmic derivative $\alpha=-(d\ln F_\nu/d\ln t)$ of the 
        afterglow synchrotron flux at frequencies below ({\it thick lines}) and above the 
        cooling frequency ({\it thin lines}), for homogeneous ({\it left panel}) and wind-like 
        circumburst media ({\it right panel}), and for a power-law angular distribution
        of the kinetic energy per solid angle of the ejecta $\cal E$ with angle $\theta$ 
        off the symmetry axis of the fireball -- ${\cal E} (\theta) \propto \theta^{-q}$ --
        outside a core of uniform density. Solid curves are for a uniform jet with sharp 
        boundaries (corresponding to $q=\infty$) and undergoing sideways expansion. 
        The observer is placed on the fireball axis, time being measured in deceleration 
        timescales for the ejecta moving along the fireball axis. When the fireball has
        decelerated enough that the edge of its core becomes visible, $\alpha$ increases 
        and the afterglow light-curve exhibits a break.
        The electron distribution is assumed to be a power-law of exponent $-p$ with $p=2.5$. 
        The asymptotic indices expected at early and late times are indicated on the abscissa.  
        Note the faster transition between the asymptotic indices in the case of a homogeneous 
        medium and that, for $q \geq 2$, the post-break index is independent of $q$, as the 
        afterglow emission arises mostly from the fireball core.}
\end{figure*}

 Figure 1 also shows the light-curve steepening for a fireball with $q=\infty$, corresponding to 
a uniform, collimated outflow with a sharp edge, undergoing lateral spreading. Due to the widening 
of the jet aperture, $\alpha (t < t_c)$ is not constant, but increases slowly in time. Furthermore, 
$\alpha (t \gg t_c)$ is larger than without lateral spreading, reaching $\alpha_2 = p$ (Rhoads 1999). 
For a uniform, spreading jet, interacting with a homogeneous medium, the fastest 70\% of the 
analytically expected steepening $\Delta \alpha = p - \alpha_1$ is acquired over a factor 20 in 
time if $\nu < \nu_c$ and 10 if $\nu_c < \nu$, while for a wind-like medium the corresponding 
factors are 100 and 1,000, respectively.
Therefore the "jet-break" transition is significantly slower and smoother if the surrounding medium 
is the wind that a massive GRB progenitor expelled before the burst (Kumar \& Panaitescu 2000).

\vspace*{3mm}
\subsection{Observers outside the Fireball Core}
\label{offcore}
\vspace*{2mm}

 For observers located outside the uniform fireball core, the pre-break light-curve fall-off 
is mitigated as the more energetic ejecta located closer to the fireball axis become visible, 
therefore the early time light-curve index $\alpha_1$ decreases in time (see also the light-curves 
presented by Granot \& Kumar 2003 and Wei \& Jin 2003). The fireball axis is seen at a time 
satisfying
\begin{equation}
 t_a \propto [{\cal E}_0 \theta_{obs}^{2(4-s)}]^{1/(3-s)} \;,
\end{equation}
similar to the time $t_c$ when the core edge becomes visible to an on-axis observer 
(\eq [\ref{tc}]) but with the core size $\theta_c$ replaced by the angle $\theta_{obs}$
between the fireball axis and the center -- observer line (Rossi, Lazzati \& Rees 2002). 
Well after $t_a$, when the beaming factor ${\cal D}$ is almost the same across the fireball 
core, the light-curve index for an off-core observer should reach that given in equations 
(\ref{a20})--(\ref{a2}) for an on-axis observer. Thus, the flattening of the light-curve 
fall-off at $t< t_a$  yields larger light-curve breaks $\Delta \alpha$ than for an on-axis 
observer. This is the most important feature arising from the structure of the outflow. 

 Figure 2 shows the evolution of $\alpha$ obtained from equation (\ref{fnu2}) for an 
off-core observer located at $\theta_{obs} = 3 \theta_c$. As can be noticed, a stronger 
fall-off of the ejecta energy away from the fireball axis (\ie a larger parameter $q$), 
leads to a more prominent flattening of the afterglow light-curve at $t < t_a$, where 
$t_a \sim 3 \times 10^3\, t_d (\theta=0)$ for $s=0$ and $\sim 10^5\, t_d (\theta=0)$ 
for $s=2$. Also as expected, at $t \gg t_a$ the index $\alpha$ asymptotically reaches 
the values for an on-axis observer. Note that, for the same parameter $q$, the light-curve 
break $\Delta \alpha$ across $t_a$ is larger for a homogeneous medium, and that the 
transition between the lowest value of $\alpha$ and the $\alpha_2$ at late time takes 
about a decade in time for a homogeneous medium and about two decades for a wind-like 
medium. The slow transition in the latter case (see also Dai \& Gou 2001 and Granot \& 
Kumar 2003) suggests that, for a wind-like stratified medium, light-curve breaks arising 
from the structure of the outflow may be too shallow compared the light-curve steepenings 
observed in some afterglows, which last less than a decade in time.

\begin{figure*}
\centerline{\psfig{figure=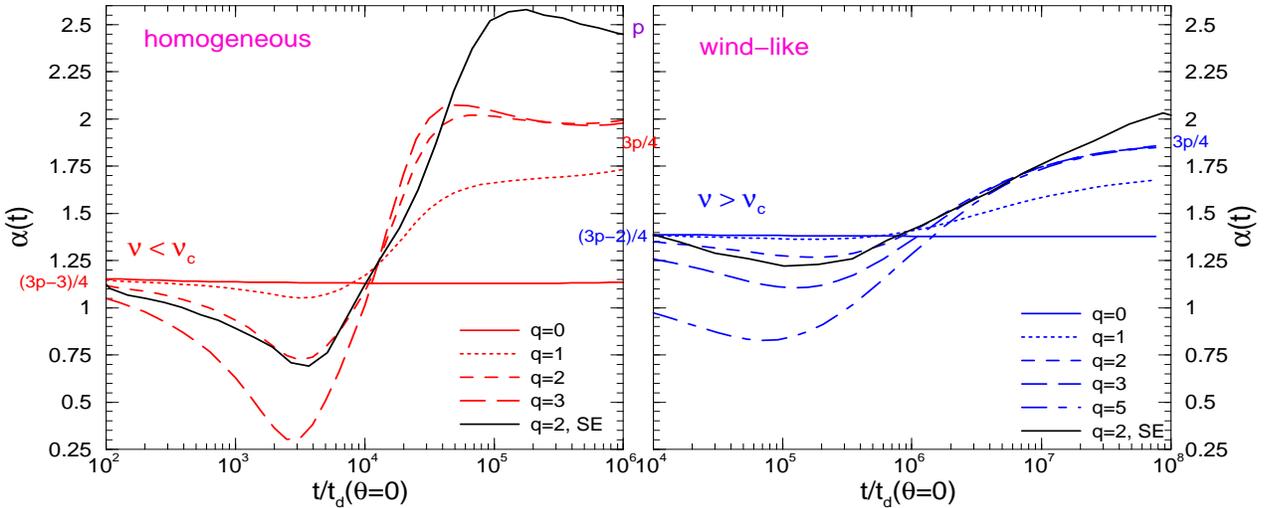,width=17cm,height=7cm}}
\figcaption{Evolution of the light-curve index $\alpha$ for various parameters $q$ for the 
       power-law fireball structure and for an observer located at angle $\theta_{obs} = 3\, 
       \theta_c$ relative to the fireball symmetry axis, $\theta_c$ being the angular size 
       of the uniform core. The electron distribution parameter is $p=2.5$. 
       When the fireball core becomes visible, $\alpha$ has a minimal value which depends 
       on $q$ and on the ratio $\theta_{obs}/\theta_c$ (Figure 3). 
       Note the slower evolution of $\alpha$ for a wind-like medium ({\it right panel}) and 
       the weaker effect that the structure has on $\alpha$ in this case. 
       For the remaining cases ($\nu > \nu_c$ and homogeneous medium, $\nu < \nu_c$ and 
       wind-like medium), the evolution of $\alpha$ is similar, the curves being shifted 
       upward by 1/4. 
       The continuous black line is for a jet of initial opening $\Theta_0 = 3 \theta_c$,
       (\ie  the observer is located on the jet edge at $t=0$) and for $q=2$.}
\end{figure*}

 The minimal value $\alpha_{min}$ reached by the light-curve index before the break depends 
not only on the structural parameter $q$, as shown in Figure 2, but also on the location 
of the observer, through the ratio $\theta_{obs}/\theta_c$, and on the slope $p$ of the 
electron distribution. As illustrated in Figure 3, the observer 
location has a much stronger effect on the sharpness of the light-curve break produced
by the fireball structure if the external medium is homogeneous. In this case, for
observer directions further away from the fireball axis, $\alpha_{min}$ decreases and
the transition between $\alpha_{min}$ and the $\alpha_2$ at late times lasts shorter
(relative to the break-time $t_a$), the light-curve break becoming sharper. Consequently, 
for a homogeneous medium, higher observer offsets will accommodate easier some of the 
observed sharp breaks. However, given the expected correlation between the GRB peak flux 
and the energy of ejecta moving toward the observer, large observer offsets render the 
burst less likely to be detected. Furthermore, for homogeneous media, as shown in Figure 3, 
large offsets also yield pre-break light-curve indices that are too small compared to
those observed. Taking $\theta_{obs}/\theta_c = 3$ as a more likely case, so that the 
resulting GRB is sufficiently bright and the afterglow break sufficiently sharp, we show 
in Figure 4 the dependence of the pre-break minimum light-curve index $\alpha_{min}$ on
the structural parameter $q$ and on the electron index $p$. Because most of the effect
of the latter is through the $3p/4$ factor in equation (\ref{a1}), the $\alpha_{min}(q)$ 
has been offset by its value $\alpha_{min}(q=0)$ for a uniform fireball, to illustrate 
the less trivial effect of $p$ on $\alpha_{min}$.

\begin{figure*}
\centerline{\psfig{figure=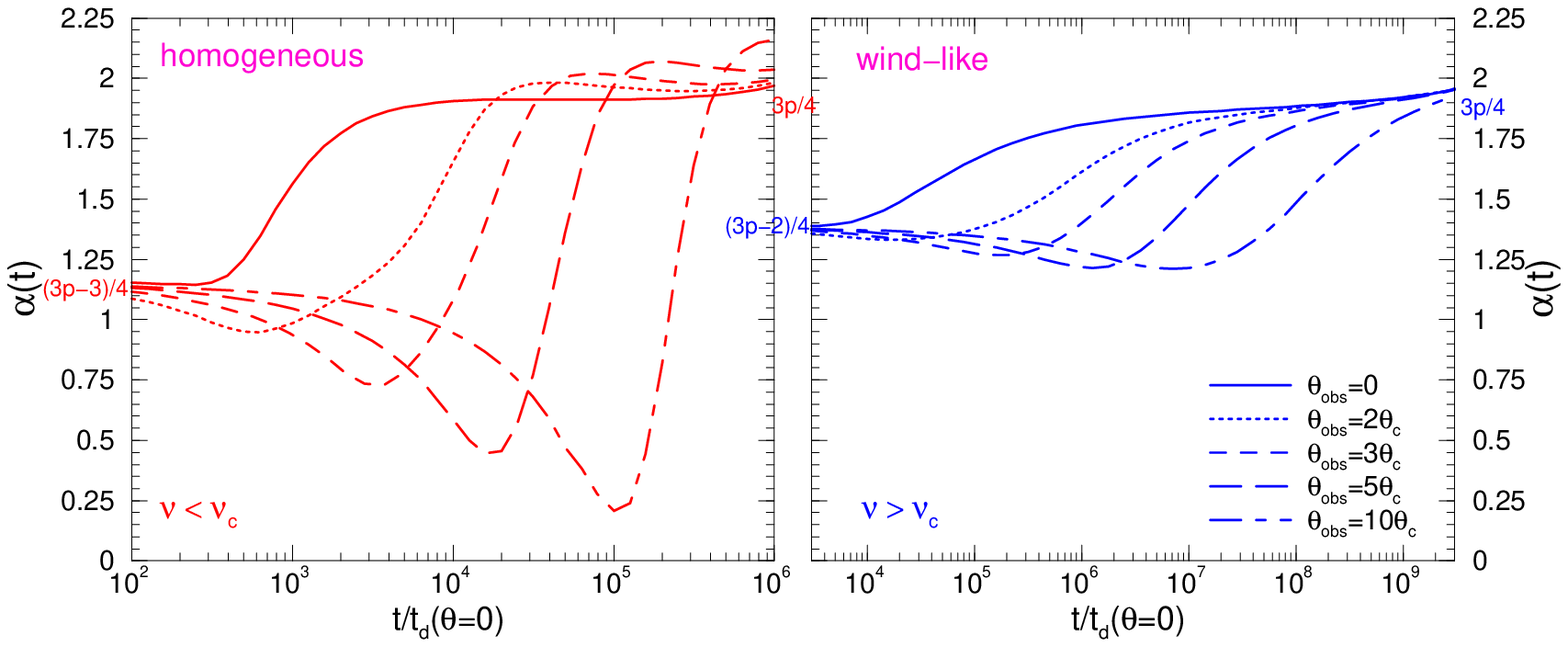}}
\figcaption{Same as in Figure 2 but for a fixed parameter $q=2$ and various ratios 
       $\theta_{obs}/\theta_c$. The line coding is the same for both panels. Note that 
       observers located further away from the fireball symmetry axis will see a stronger 
       flattening of the afterglow light-curve followed by a sharper break, and that the 
       break is shallower for a wind-like medium ({\it right panel}).}
\end{figure*}

\begin{figure*}
\centerline{\psfig{figure=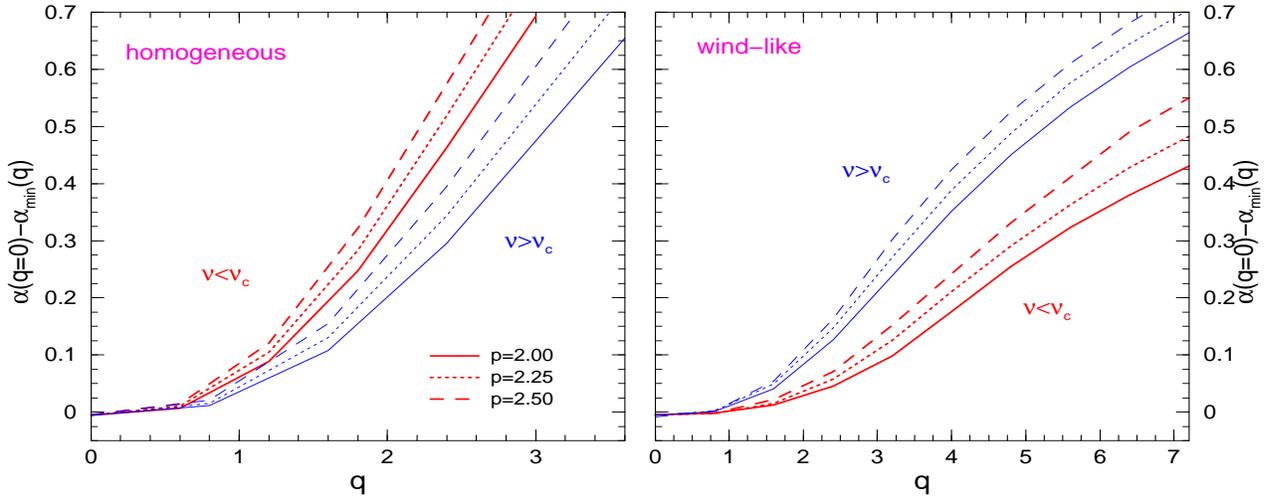,width=17cm,height=7cm}}
\figcaption{Effect of fireball structure (parameterized by $q$) on the minimum light-curve 
       index $\alpha_{min}$ reached before the break, for some values of the electron index 
       $p$.  The observer location is $\theta_{obs} = 3 \theta_c$. {\it Thick curves} are for 
       $\nu < \nu_c$ while {\it thin lines} indicate $\nu > \nu_c$. Most of the dependence 
       of $\alpha_{min}$ on $p$ has been eliminated by shifting vertically the curves by the 
       $\alpha (q=0)$ expected for a uniform fireball. Note the stronger flattening (\ie 
       decrease of $\alpha_{min}$) that a homogeneous medium ({\it left panel}) yields for 
       the same structural parameter $q$.}
\end{figure*}

\vspace*{3mm}
\section{Structured Fireballs/Jets and Afterglow Observations} 
\vspace*{2mm}

 The minimum light-curve index $\alpha_{min}$ shown in Figure 2 should be close to the 
pre-break index usually inferred by fitting the afterglow light-curve with a smoothed 
broken power-law (\eg Beuermann \etal 1999), while the post-break index inferred from 
observations is either the $\alpha_2$ given in equations (\ref{a20})--(\ref{a2}) for a 
structured fireball, or the electron index $p$ for a jet (uniform or with structure) 
whose edge is visible. Furthermore, one can determine from multiband optical observations 
the afterglow intrinsic spectral slope $\beta_o$, defined as $F_\nu \propto \nu^{-\beta_o}$, 
which depends only on the electron index $p$:
\begin{equation}
 \beta_o = \left\{\hspace*{-2mm} \begin {array}{ll} 
               (p-1)/2  & \nu_i < \nu_o < \nu_c \\ 
                p/2     & \max\{\nu_i,\nu_c\} < \nu_o  
  \end{array} \right. \;.
\label{beta}
\end{equation}
If there is a significant dust reddening in the host galaxy, $\beta_o$ can be determined
with the aid of the $X$-ray spectral slope, if available, or of fits to the optical spectrum,
if it is sufficiently curved to constrain the host extinction for an assumed dust-reddening 
law (\eg Jensen \etal 2001, Fynbo \etal 2001). Thus, observations provide three characteristics 
of the afterglow emission: pre- and post- break light-curve indices $\alpha_{min}$ and 
$\alpha_2$, and spectral slope $\beta_o$, which constrain three major fireball parameters: 
the structural index $q$, the observer location $\theta_{obs}/\theta_c$, and the electron 
index $p$. 

 An analytical determination of $q$ from the measured
asymptotic light-curve indices is hampered by the complicated dependence of the pre-break
index $\alpha_{min}$ on $q$ and observer location (Figures 2 and 3) and by that 
the post-break index $\alpha_2$ could be affected by the lateral spreading of the outflow. 
The observer location $\theta_{obs} = 3\,\theta_c$ considered in Figures 2 and 4 
can be taken as representative, but, for a more secure approach to determining the fireball 
structure, one should fit the observed break with a numerically calculated afterglow 
light-curve, allowing for a free observer location. Before proceeding on this path, we
discuss below how the observed asymptotic light-curve indices $\alpha_1$ and $\alpha_2$
and the spectral slope $\beta_o$ can be used to assess if a structured fireball is required
by the observations and if the lateral spreading of the outflow is significant.

\vspace*{2mm}
\subsection{How to Constrain the Outflow Collimation and Structure Using Observations}
\label{criteria}
\vspace*{2mm}
 
 We make here the {\it assumption} that the asymptotic pre- and post-break light-curve indices
are determined only by the fireball/jet angular structure, radial distribution of the external
medium density, and slope of the shock-accelerated electron distribution. In principle there could
be other factors that determine these indices, such as: a refreshed shock produced by slower 
ejecta catching-up with the GRB remnant (Panaitescu, M\'esz\'aros \& Rees 1998), which may 
mitigate the afterglow dimming rate if sufficient energy is injected, or a dependence on the 
ejecta Lorentz factor of the electron distribution index or fractional energy acquired by 
electrons or magnetic field.

 From equations (\ref{a1}) and (\ref{beta}) it follows that, for a uniform outflow, the pre-break 
light-curve index is
\begin{equation}
 \alpha_1 = \frac{3}{2} \beta_o + \left\{  \hspace*{-2mm} \begin{array}{ll} 
        0   & \;\; s=0   \;\; \& \;\; \nu_i < \nu_o < \nu_c \\ 
       -1/2 & \;\; s=0,2 \;\; \& \;\; \max\{\nu_i,\nu_c\} < \nu_o \\ 
        1/2 & \;\; s=2   \;\; \& \;\; \nu_i < \nu_o < \nu_c 
        \end{array} \right. \;.
\label{ab1}
\end{equation}
Thus, {\it if the observed $\alpha_1$ is smaller than given above, a structured outflow is required}
($1^{st}$ criterion), as well as an off-core observer location. Given that for a fireball with a 
decreasing $\cal E (\theta)$, the post-break light-curve index $\alpha_2$ cannot exceed the values 
given in equation (\ref{a2}), {\it if the observed $\alpha_2$ exceeds}  
\begin{equation}
 \alpha_2 = \frac{3}{2} \beta_o + \left\{  \hspace*{-2mm} \begin{array}{ll}
            3/4  &  s=0  \;\; \& \;\;  \nu_i < \nu_o < \nu_c \\
            1/4  &  s=0  \;\; \& \;\;  \max\{\nu_i,\nu_c\} < \nu_o \\
             1   &  s=2  \;\; \& \;\;  \nu_i < \nu_o < \nu_c \\ 
             0   &  s=2  \;\; \& \;\;  \max\{\nu_i,\nu_c\} < \nu_o 
          \end{array} \right. \;,
\label{ab2}
\end{equation}
{\it then a highly collimated outflow (jet)}, whose boundary becomes visible when the light-curve 
break is observed, {\it is required} ($2^{nd}$ criterion). In this case, the post-break light-curve 
index reaches its maximal value: $\alpha_{max} = p$. From equation (\ref{beta}), it follows that
\begin{equation}
 \alpha_{max} = \left\{  \hspace*{-2mm} \begin{array}{ll} 
            2\beta_o + 1  &  \nu_i < \nu_o < \nu_c \\ 
            2 \beta_o     &  \max\{\nu_i,\nu_c\} < \nu_o  
             \end{array} \right. \;.
\label{alfamax}
\end{equation}
Combining equations (\ref{ab1}) and (\ref{alfamax}), the maximum steepening $\Delta \alpha = 
\alpha_2 - \alpha_1$ that a uniform jet can produce is 
\begin{equation}
 \Delta \alpha =  \frac{1}{2} \beta_o + \left\{  \hspace*{-2mm} \begin{array}{ll} 
        1    &  s=0 \;\; \& \;\; \nu_i < \nu_o < \nu_c \\ 
        1/2  & (s=0 \;\&\; \max\{\nu_i,\nu_c\} < \nu_o) \;\; {\rm or} \;\; s=2 
        \end{array} \right. \;.
\label{Da}
\end{equation}
Thus, {\it if the observed $\Delta \alpha$ exceeds the above value, a structured jet is required}
($3^{rd}$ criterion). However, {\it if the observed $\alpha_2$ exceeds that given in equation 
(\ref{alfamax})}, then the light-curve break cannot be due entirely to the collimation of the 
ejecta, and {\it another mechanism for light-curve breaks}, such as a spectral break passing 
through the observing band, {\it is required} ($4^{th}$ criterion). This passage will also 
produce a color change (\ie $\beta_o$ increases) and an achromatic light-curve break. 

 Applying the above criteria to the optical emission of afterglows with light-curve breaks, 
it is possible to identify the cases where structured fireballs or jets are required: \\
$i)$ from the $1^{st}$ criterion, angular structured is required by the pre-break index 
     $\alpha_1$ of the afterglows \\ 
   \hspace*{3mm}  $a)$ 980519, 990510, and 991216, if $s=2$,  \\
   \hspace*{3mm}  $b)$ 000301c and 010222, if $\nu_o < \nu_c$, \\
   \hspace*{3mm}  $c)$ 990123 and 000926, if $s=2$ and $\nu_o < \nu_c$, \\
$ii)$ from the $2^{nd}$ criterion, a jet undergoing lateral spreading is required by the
     post-break index $\alpha_2$ of the afterglows \\
   \hspace*{3mm}  $a)$ 990510, \\
   \hspace*{3mm}  $b)$ 990123, 000926, and 010222, if $\nu_c < \nu_o$, \\
   \hspace*{3mm}  $c)$ 991208 if $\nu_o < \nu_c$, \\
$iii)$ from the $3^{rd}$ criterion, a structured jet is required by the steepening $\Delta \alpha$
      observed for the afterglow 990510, if $s=2$, \\ 
$iv)$ from the $4^{th}$ criterion, the passage of a spectral feature through the optical domain
      is required (Li \& Chevalier 2001, Panaitescu \& Kumar 2002) by the steep post-break index 
      $\alpha_2$ of the afterglows 991208 and 000301c, if $\nu_c < \nu_o$ and, perhaps, by 000301c 
      if $\nu_o < \nu_c$ .

\vspace*{2mm}
\subsection{Numerical Modeling of Structured Jets}
\label{models}
\vspace*{2mm}

 Equations (\ref{ab1})--(\ref{Da}) allow one to assess if a structured outflow is necessary
or sufficient to explain the pre and post-break optical light-curve indices. A more conclusive 
test should investigate if the light-curve steepening produced by a structured fireball is as
sharp as observed, and if the fireball emission at different wavelengths is consistent with 
the observations. For uniform outflows, the latter is usually done using a snapshot broadband 
afterglow spectrum, but, for structured outflows, one has to take in account that, at the same
observer time, the ejecta moving at different angles have different Lorentz factors, thus their
synchrotron emissions have with different characteristic frequencies.  

 To perform such a test, the data should be fit numerically, followed by a comparison
of the best fit $\chi^2$ obtained for the various model features that are investigated. 
The model used hereafter takes into account the spread in the photon arrival time and Doppler 
boost due to the curvature of the jet surface, radiative losses, the effect of inverse 
Compton scatterings on the cooling frequency, and the self-absorption and interstellar
scintillation at radio frequencies. The uniform jet expands sideways at a comoving frame 
tangential velocity equal to the sound speed $c_s$. Thus, the jet opening $\Theta$ is given 
by 
\begin{equation}
 d \Theta (r) = \frac{c_s}{r} dt' = \frac{c_s}{r \gamma}\, dr\;,  
\label{Theta}
\end{equation}
where $t'$ is time measured in the comoving frame. The uniformity of the jet is assumed to 
be maintained during the jet lateral spreading. The variation of the energy per solid angle 
${\cal E}$ of an infinitesimal ring $(\theta, \theta + \delta \theta)$ during $dt'$ is 
$d {\cal E} = - {\cal E}\, d(\delta \theta)/ \delta \theta$, where $d(\delta \theta)$ is 
the spreading of the ring during $dt'$. Therefore, the assumption of jet uniformity at any 
time is equivalent to that $d(\delta \theta)/ \delta \theta$ is a $\theta$-independent quantity. 
Then, from equation (\ref{Theta}), it follows that the spreading of any infinitesimal 
ring is assumed to be governed by
\begin{equation}
 d(\delta\theta) =  \frac{\delta\theta}{\Theta} \frac{c_s}{r \gamma}\, dr \;. 
\label{deltheta}
\end{equation}

 The collimation of ejecta affects the afterglow emission in three ways. First, the size of
the visible part of the jet ($r/\gamma$) stops increasing when $\gamma \Theta = 1$, \ie when 
the edge is seen, yielding a light-curve steepening (Panaitescu, M\'esz\'aros \& Rees 1998). 
Second, the sideways expansion of the jet increases its sweeping area, which alters the jet 
dynamics (Rhoads 1999). Third, the lateral spreading decreases the surface density of the 
swept-up medium, moving a fraction of the radiating electrons outside the visible region 
when the jet edge is not yet seen. The last two effects are important when $\Theta - \Theta_0 
\simg \Theta_0$ ($\Theta_0$ being the initial jet opening), which, with the aid of equations 
(\ref{gm}) and (\ref{Theta}), can be shown to occur slightly {\it after} the radius where the 
jet edge becomes visible if the external medium is homogeneous, and slightly {\it before} if 
the medium is wind-like. The jet lateral spreading contributes to the light-curve steepening 
caused by seeing the jet edge, but also delays the completion of the steepening. Because
of the slower deceleration caused by a wind-like medium, this completion lasts longer for 
$s=2$ than for $s=0$. Therefore, the light-curve breaks for spreading jets should be smoother 
than for structured fireballs, the effect being stronger for a wind-like medium than for 
a homogeneous one (Figures 1 and 2). 

 For a structured fireball, we divide its surface in $(\theta_i, \theta_i + \delta \theta_i)$ 
rings and calculate the ring location ($\theta_i (r)$) and width ($\delta \theta_i (r)$)
using equation (\ref{deltheta}), with the local sound speed $c_s (\theta_i)$ obtained from the 
shock jump conditions and dynamics ($\gamma_i (r)$) of the ring, and with $\Theta (r) = \sum_i 
\delta \theta_i (r)$. For each step $r \rightarrow r + dr$, the deceleration of the ring is 
calculated as if it were part of a uniform fireball of kinetic energy per solid angle ${\cal E} 
(\theta_i)$, which is adjusted at each step to account for the ring spreading $\delta \theta_i 
(r)$ and radiative losses. Therefore, we ignore any fluid mixing expected to arise from the 
tangential flow, a factor which Granot \& Kumar (2003) have found to have a weak effect on the 
afterglow emission. Equation (\ref{deltheta}) provides a simple one-to-one mapping of the jet 
structure at radius $r$ into that at $r+dr$, and maintains computationally inexpensive the 
numerical algorithm for the dynamics. 

 The uniform jet model has six free parameters: the initial energy density per solid angle within
the ejecta $\cal E$, the jet initial half-opening $\Theta_0$, the external particle density 
$n$ (or the parameter $A_*$ introduced by Chevalier \& Li 1999 for an $r^{-2}$ wind profile),
the exponent $p$ of the electron distribution, and the fractional energies in electrons and 
magnetic field in the post-shock gas. For observers located within the initial jet opening
(which is a condition for seeing a GRB), this location is irrelevant (Granot \etal 2002). 
The structured jet model with a power-law distribution of the energy per solid angle has three 
additional parameters: the core size $\theta_c$, the power-law exponent $q$, and the observer 
location $\theta_{obs}$. 

 We apply the structured jet model to the two afterglows with the largest observed steepening
$\Delta \alpha$: 990510 and 000301c. For 990510, a jet undergoing lateral spreading at the
time of observations is required for either type of external medium. The purpose of modeling
this afterglow is twofold: for $s=0$ we will determine how much structure is allowed by the 
pre-break light-curve index, for $s=2$ we will assess if a structured outflow improves significantly 
the best fit by sharpening the light-curve steepening. The latter is also the purpose of modeling 
the afterglow 000301c with a structured jet, as its sharp light-curve break cannot reproduced with 
a uniform jet and standard electron distribution even if the external medium is homogeneous.

\subsubsection{GRB 990510}
\label{0510}

 The optical emission of the 990510 afterglow exhibited a break from an index $\alpha_1 = 0.82 
\pm 0.02$ to $\alpha_2 = 2.18 \pm 0.05$ (Harrison \etal 1999) at $t \sim 2$ days (see Figure 6). 
The optical and $X$-ray slopes, $\beta_o = 0.61 \pm 0.12$ (Stanek \etal 1999) and $\beta_x = 1.03
\pm 0.08$ (Kuulkers \etal 2000), indicate that $\nu_o < \nu_c < \nu_x$.

 In the framework of uniform jets, we have previously found (Panaitescu \& Kumar 2002) that
the optical light-curve steepening can be well accommodated  ($\chi^2 = 36$ for 69 degrees 
of freedom) with a jet of initial kinetic energy $E_0 = (1 \div 6) \times 10^{50}$ ergs, initial
opening $\Theta_0 \simeq 3 \deg$, interacting with a homogeneous medium of density $n = 0.1 
\div 0.4\, \cm3$. We have also found that a wind-like medium cannot reproduce the steepness 
of the optical light-curve break, yielding an unacceptable best fit ($\chi^2 = 127$ for 69 df). 

 For a {\it homogeneous} external medium, equation (\ref{ab1}) shows that $\alpha_1 = 0.92 \pm
 0.18$ for a uniform outflow, which is consistent with the observed index, thus a large structural 
parameter $q$ is disfavored, otherwise the early decay would be too shallow. Further, equation 
(\ref{ab2}) shows that, for a fireball, the post-break index $\alpha_2$ cannot exceed $1.67 \pm 
0.18$, which is too small compared to that observed, thus a spreading jet is required. In this 
case, equation (\ref{alfamax}) leads to a post-break index $\alpha_2 = 2.22 \pm 0.24$, which is 
consistent with the measured index.
The above conclusion that the pre-break light-curve index of the afterglow of GRB 990510
does not allow much structure in a spreading jet is illustrated in Figure 5, showing the
$\chi^2$ of the fits obtained for several combinations of the structural parameter $q$ and 
jet core size-to-opening ratio $\theta_c/\Theta_0$, and for two particular observer locations:
the jet axis and the jet edge. By decreasing the $\theta_c/\Theta_0$ ratio or increasing $q$, 
a stronger jet structure is enforced. As can be noticed, only jets with $\theta_c \simg 0.2\, 
\Theta_0$ for $q=1$ and $\theta_c \simg 0.3\, \Theta_0$ for $q \geq 2$ provide acceptable fits 
(defined by a probability larger than 10\%), while fits as good as that provided by a uniform 
jet ($\chi^2 = 36$) require that $\theta_c \simg 0.3\,\Theta_0$ if $q=1$ and $\theta_c \simg 
0.4\,\Theta_0$ if $q=2$. Thus, for $s=0$, the afterglow of GRB 990510 is best explained by a 
structured jet with a variation of the energy per solid angle less than a factor $\sim 5$ 
across the jet surface. The radio, optical and $X$-ray light-curves for the best fit obtained 
with a structured jet are shown in the left panel of Figure 6. We note that its parameters 
are similar to those for a uniform jet.

\begin{figure*}
\centerline{\psfig{figure=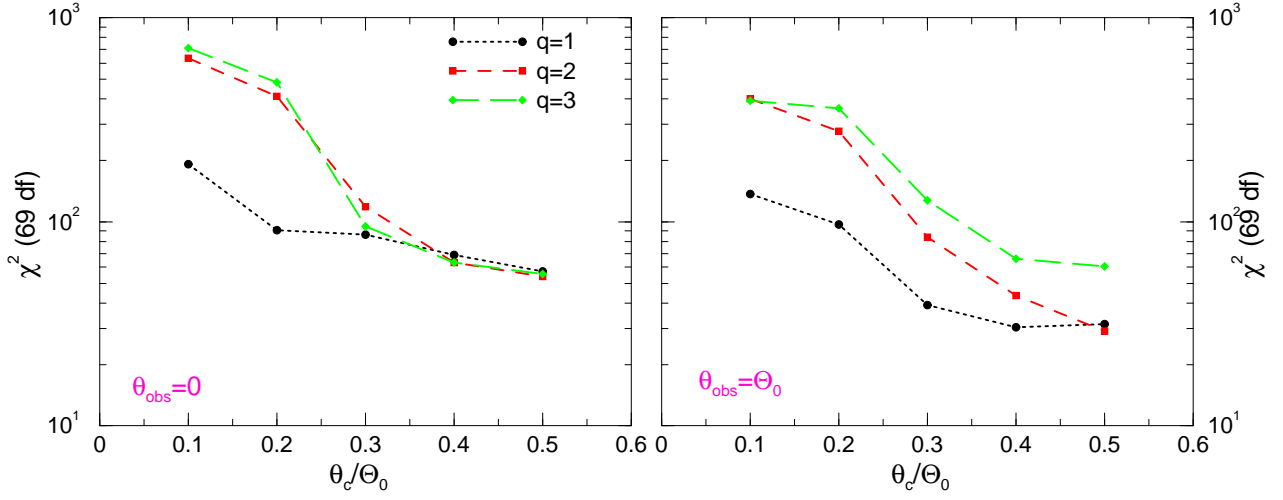,width=17cm,height=7cm}}
\figcaption{Variation of the $\chi^2$ of the best fits to the radio, optical, and $X$-ray
      emission of the 990510 afterglow (69 degrees of freedom) with the amount of structure 
      in the spreading jet interacting with a {\it homogeneous medium}. 
      {\it Left panel}: observer on jet axis, {\it right panel}: observer on jet edge. 
      Each point represents a combination of the ratio $\theta_c / \Theta_0$ and structural
      parameter $q$. Other model parameters (\S\ref{models}) were left free.
      Note that, as the angular distribution of the jet energy is more anisotropic, the fits  
      become poorer. The best fit obtained with a uniform jet has $\chi^2 = 36$. Fits of 
      comparable quality are obtained only if the variation of the energy per solid angle 
      across the jet surface is less than a factor $\sim 5$.}
\end{figure*}

\begin{figure*}
\centerline{\psfig{figure=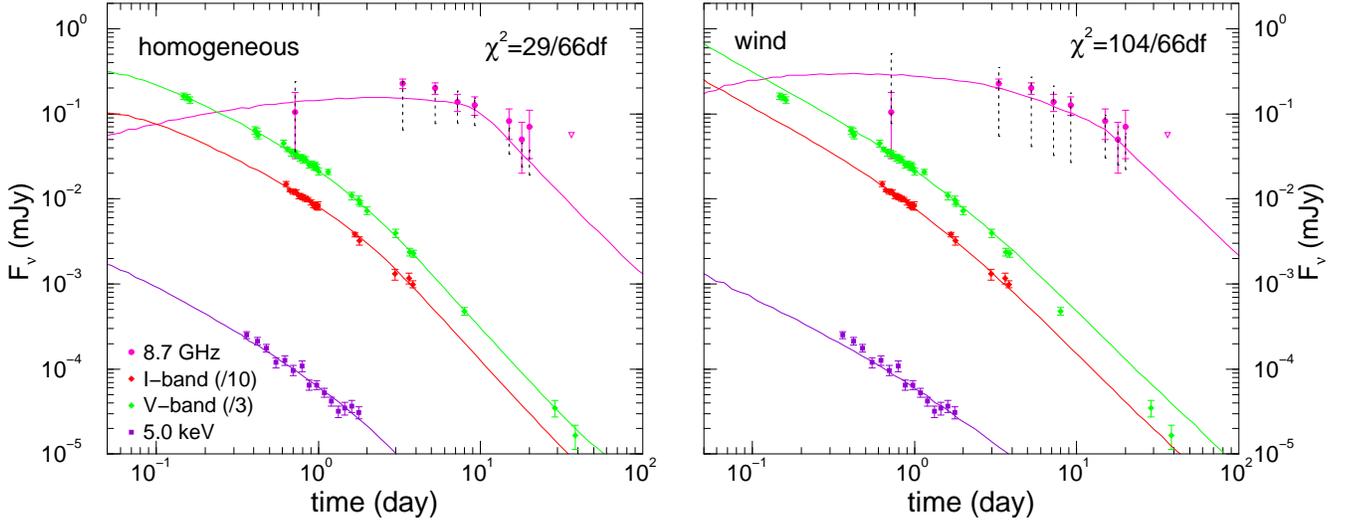}}
\figcaption{Best fits for the 990510 afterglow obtained with structured jets. 
      {\it Left panel} -- jet parameters are: initial energy density per solid angle along
      the jet axis ${\cal E}_0 \simeq 10^{53}$ ergs/sr, initial jet half-opening $\Theta_0 
      \simeq 2.2\, \theta_c$, observer location $\theta_{obs} \simeq 2.2\,\theta_c$, size of 
      jet uniform core $\theta_c \simeq 0.8\deg$, exponent of power-law angular distribution 
      of energy within the jet $q \simeq 1.5$ (thus the initial energy of the jet is $E_0 
      \simeq 2 \times 10^{50}$ ergs), electron index $p \simeq 1.8$, and {\it homogeneous 
      medium} of particle density $n \simeq 0.2\, \cm3$, yielding $\chi^2 = 29$ for 66 degrees 
      of freedom (df). 
      {\it Right panel} -- ${\cal E}_0 \simeq 8\times 10^{52}$ ergs/sr, $\Theta_0 \simeq 
      4.6\,\theta_c $, $\theta_{obs} \simeq 1.8\, \theta_c$, $\theta_c \simeq 1 \deg$, 
      $q = 2 \div 3$ (therefore $E_0 \siml 3 \times 10^{50}$ ergs), $p \simeq 1.8$, and 
      {\it wind-like medium} with $A_* \simeq 0.2$, yielding $\chi^2 = 104$ for 66 df. 
      In both cases, the cooling frequency falls in the optical range, the spectral slope 
      being in between the values given in equation (\ref{beta}). 
      The vertical dotted lines indicate the amplitude of the fluctuations at 9 GHz caused 
      by inhomogeneities in the Galactic interstellar medium. Triangles indicate $2\,\sigma$
      upper limits on the 9 GHz emission.
      For clarity, the optical fluxes have been shifted by the factors indicated in the legend. 
      Note that, for a wind-like medium, the steepening of the light-curve is too slow, failing 
      to accommodate the sharpness of the break seen at 2-3 days.}
\end{figure*}

 For a {\it wind-like} external medium, equation (\ref{ab1}) gives $\alpha_1 = 1.42 \pm 0.18$
for a uniform outflow, thus structure is required in this case to explain the slower decay 
observed at early time. Equation (\ref{ab2}) shows that a structured fireball yields 
$\alpha_2 \leq 1.92 \pm 0.18$, slightly smaller than the observed index, suggesting that 
collimation may also be required to accommodate the post-break decay of the this afterglow.
The right panel of Figure 6 shows the best fit obtained with a structured jet interacting
with a wind medium. The jet structure improves the fit by $\Delta \chi^2 = 23$ relative to 
the best fit obtained with a uniform jet, which is statistically significant. However, 
the optical light-curves steepen too slowly, overestimating the afterglow flux before 
and after the break and yielding $\chi^2 = 104$ for 66 df. We conclude that a wind-like 
medium is not compatible with the observations of the 990510 afterglow even if the jet is
endowed with structure.

\subsubsection {GRB 000301c}
\label{0301}

 The $R$-band light-curve index of the 000301c afterglow increased from $\alpha_1 = 0.70 
\pm 0.07$ to $\alpha_2 = 2.44 \pm 0.29$ (Bhargavi \& Cowsik 2000), a sharp break occurring
at $t \sim 4$ days (Figure 7). Jensen \etal (2001) found that, at $t = 3$ days,
the optical spectral slope is $\beta_o = 0.57 \pm 0.02$ after correcting for the host 
reddening, determined from the curvature of the spectrum and assuming an SMC reddening law. 
$X$-ray observations were not made, thus the location of the cooling frequency is not 
constrained. If $\nu_c < \nu_o$ then the pre-break light-curve index (\eq[\ref{ab1}]) 
$\alpha_1 = 0.36 \pm 0.03$ for a uniform outflow would be too small compared with that 
observed. Therefore $\nu_o < \nu_c$ seems more likely for the 000301c afterglow.

 Given that structured jets interacting with a wind-like medium produce light-curves that
are too smooth, we focus here only on a homogeneous circumburst medium. For $\nu_o < \nu_c$ 
and a uniform fireball, equation (\ref{ab1}) yields $\alpha_1 = 0.86 \pm 0.03$, which is 
slightly larger than observed, thus a structured outflow is only marginally required,
while equation (\ref{ab2}) leads to $\alpha_2 \leq 1.61 \pm 0.03$, which is well below the 
observed value, thus a jet is required. Then, according to equation (\ref{alfamax}), 
$\alpha_2 = 2.14 \pm 0.04$, which is marginally compatible with the index measured
by Bhargavi \& Cowsik (2000). Other post-break asymptotic indices reported in the 
literature are larger (albeit more uncertain), which suggests that a jet model may have 
difficulties in explaining the steep post-break fall-off of the $R$-band light-curve of 
000301c.

 In the framework of uniform jets, we have found (Panaitescu 2001) that this indeed the case: 
the best fit obtained for $s=0$ has $\chi^2 = 480$ for 98 df and $E_0 \simeq 2 \times 10^{51}$ 
ergs, $\Theta_0 \simeq 3 \deg$, $n \simeq 0.01\, \cm3$, $p \simeq 2.5$, failing to produce 
the observed steepening $\Delta \alpha \geq 1.74 \pm 0.30$ when the jet edge becomes visible. 
For this reason, we have investigated a jet model where the distribution of shock-accelerated 
electrons has a break, so that a sharp light-curve fall-off is seen after the passage through 
the optical of the synchrotron characteristic frequency corresponding to this break. 
Further indication that the distribution of injected electrons is not a pure power-law is 
provided by the discrepancy between the post-break light-curve indices at radio ($\alpha_r 
= 1.0 \pm 0.2$) and optical frequencies, and also by the $K-R$ color change between 2 and 5 
days after the burst (Rhoads \& Fruchter 2001), which implies a softening of the optical 
spectral slope $\Delta \beta_o \siml 0.5$ that is too fast to be attributed to the passage 
of the cooling break. The best fit obtained with the broken power-law injected electron 
distribution has $\chi^2 = 120$ for 96 df, being marginally acceptable. 

 Figure 7 assesses the ability of a structured jet to accommodate the sharp break of the
000301c afterglow without recourse to a non-standard injected electron distribution. 
The new best fit has $\chi^2 = 204$ for 95 df, which is a substantial improvement
in comparison with the uniform jet model and a pure power-law electron distribution. 
Nevertheless, the best fit obtained with a structured jet is not acceptable and, clearly, 
poorer than the uniform jet model with a broken power-law injected electron distribution. 
It does not reproduced well the steep post-break decay of the $R$-band light-curve and 
underestimates the 250 GHz emission. We note that, for the best fit shown in Figure 7, 
the cooling frequency falls within the optical domain at a few days, a feature which is 
required to explain the observed curvature of the optical spectrum (Jensen \etal 2001).

\begin{figure*}
\centerline{\psfig{figure=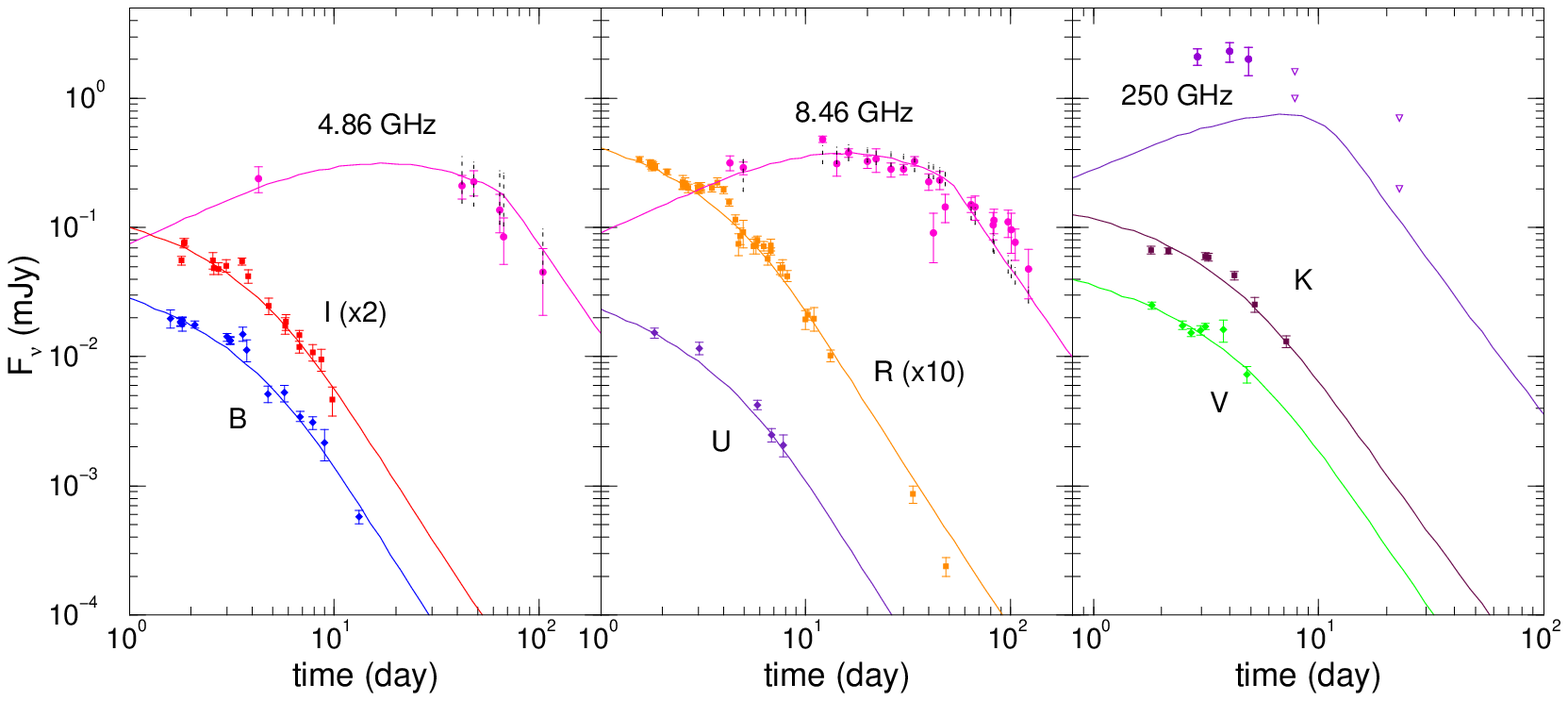}}
\figcaption{Best fit for the 000301c afterglow obtained with a structured jet and a 
      {\it homogeneous external medium}. 
      The jet parameters are ${\cal E}_0 \simeq 3\times 10^{53}$ ergs/sr, $\Theta_0 \simeq 5.0\,
      \theta_c $, $\theta_{obs} \simeq 3.3\, \theta_c$, $\theta_c \simeq 0.8 \deg$, $q \sim 3$ 
      (therefore $E_0 \sim 5 \times 10^{50}$ ergs), $p \simeq 2.5$, and $n \simeq 0.04\, \cm3$. 
      The jet energy is 4 times smaller than that obtained with a uniform jet, while the external 
      density is 4 times larger. 
      This fit has $\chi^2 = 204$ for 95 df, excluding the data between 3.0 and 4.3 days, when the 
      optical light-curves exhibit a flattening, indicating a departure (\eg delayed energy injection, 
      clumpy external medium) from the jet model used here. 
      Triangles indicate $1\,\sigma$ and $2\,\sigma$ upper limits on the 250 GHz emission. 
      As for Figure 6, vertical dotted lines give the amplitude of interstellar scintillation at 
      radio frequencies.
      Note the shallow decay of the model $R$-band emission at late times. }
\end{figure*}

\vspace*{3mm}
\section{Conclusions}
\vspace*{2mm}

 In the standard picture of uniform jets, the measured pre- and post-break light-curve 
indices and the spectral slope offer an overconstrained problem, as, for a given type of
external medium, all these quantities depend only on the exponent of the power-law distribution 
of shock-accelerated electrons. The angular structure of GRB jets and the observer location
relative to the axis of the structured jet affects the pre-break index and, possibly, the
post-break index also (\eq [\ref{a20}]--[\ref{a2}], Figures 2--4). For most well-observed 
afterglows, a uniform jet interacting with a homogeneous external medium provides good fits 
(Panaitescu \& Kumar 2002), indicating that the freedom allowed by the outflow structure is 
usually not required by current afterglow observations. 
Nevertheless, a simple analysis (\S\ref{criteria}) of the light-curve indices and spectral slopes,
shows that structured outflows are required if the circumburst medium has a wind-like profile
($s=2$), as, for such media, the resulting steep pre-break light-curve decay must be mitigated 
by the outflow angular structure.

 Due to the slower deceleration of jets by wind-like media and the stronger sideways expansion 
when the jet edge becomes visible, the light-curve breaks for $s=2$ and uniform jets are too 
smooth compared with the observations (Kumar \& Panaitescu 2000). It is then worth investigating
if an angular structure in the outflow, yielding a shallower pre-break light-curve decay, 
sharpens sufficiently the light-curve steepening to make it consistent with the observations. 
Figures 2, 3, and the right panel of Figure 6 show that the steepening of light-curves from
structured fireballs/jets interacting with wind-like media lasts more than a decade in time,
suggesting that such media are not able to accommodate all the observed afterglow breaks.

 Using numerical calculations of the dynamics and synchrotron emission of structured jets,
we find (\S\ref{0510}) that little angular structure is allowed for the afterglow 990510
if the external medium is homogeneous. Although acceptable fits to the emission of this 
afterglow can be obtained even for a tenfold variation of the kinetic energy per solid 
angle from its axis to edge, fits of quality comparable to that obtained with a uniform 
jet require an energy variation across the jet surface of less than a factor $\sim 5$. 
Allowing for structure in the jet improves significantly the best fit to 990510 obtained 
with a wind-like medium, nevertheless the best fit is not acceptable. 
The ability of structured jets to yield larger light-curve breaks was also tested against
the strong steepening observed for the afterglow 000301c (\S\ref{0301}). We find that the
addition of structure greatly improves the fit obtained with a pure power-law electron
distribution; however it does not fare as well as a jet model with a non-standard, broken
power-law electron distribution.

 Nevertheless, angular structure in a wide jet may be a viable explanation for the shallower
light-curve breaks (Rossi, Lazzati \& Rees 2002) seen in other GRB afterglows. In fact, as 
discussed in \S\ref{criteria}, a structured fireball is required by some afterglows if the 
external medium is wind-like.
Evidently, the best cases for a structured outflow will be those where the light-curve indices
and spectral slopes are not consistent with each other within the framework of uniform fireballs.
In such cases, equations (\ref{ab1})--(\ref{alfamax}) and the criteria derived in \S\ref{criteria}
offer a simple way to assess the importance of structure and collimation in the outflow.
Furthermore, equations (\ref{a1}), (\ref{a20})--(\ref{a2}), and the results shown in Figure 5
can be used to constrain the angular distribution of the ejecta kinetic energy.

\end{document}